 \documentclass[nohyper,12pt,letterpaper]{JHEP}
 \usepackage{epsfig}
 \DeclareSymbolFont{AMSa}{U}{msa}{m}{n}
 \DeclareSymbolFont{AMSb}{U}{msb}{m}{n}
 \let\Box\relax
 \DeclareMathSymbol{\Box}{\mathord}{AMSa}{"03}

 \def \eqn#1#2{\begin{equation}#2\label{#1}\end{equation}}

 \title{Cosmological Supersymmetry Breaking\\and The Power of the Pentagon:\\A Model of Low Energy
 Particle Physics}
 \author{T.\,Banks\\
 Department of Physics \\
 University of California, Santa Cruz, CA 95064\\
 E-mail: \email{banks@scipp.ucsc.edu}\\
 {\it and}\\
 NHETC, Rutgers University\\
 Piscataway, NJ 08854}

 \abstract{I present a low energy Lagrangian implementing
 the idea of Cosmological SUSY breaking (CSB).   The model predicts ${\rm tan }\beta \sim 1$, and
 incorporates a new mechanism for breaking of $SU(2)\times U(1)$. The Higgs mass is
 determined by new physics and can evade the bounds of the MSSM.
 The model resolves the CP and flavor problems of SUSY.
 The up quark mass is non-vanishing. An axion-like particle, with
TeV scale decay constant,
 appears, which could provide the solution of the strong CP problem.  Such a particle is
 experimentally ruled out if it has conventional QCD axion couplings. The problem {\it may} be
 avoided by adding dimension 5 operators, which explicitly break the axial
 symmetry. However, it is likely that this reintroduces the strong
 CP problem.}
 \keywords{Cosmological SUSY Breaking}
 \received{} \accepted{}
 \preprint{\hepph{0510159}\\\\ \\}
 
\begin{document}
 \section{\bf Introduction}

In a previous paper\cite{susycosmophenoIII}, the author proposed a
class of low energy models, which implemented the idea of
Cosmological SUSY Breaking (CSB)\cite{tbfolly}\cite{susyhor}. The
models contained an elementary singlet chiral superfield $G$, and a
new strongly coupled gauge theory ${\cal G}$, which contained matter
multiplets with standard model quantum numbers. $G$ was coupled to
singlet bilinears in the new gauge theory, as well as $H_u H_d$. The
model had an exact discrete R symmetry, with $R_G = 0$, which
forbade a superpotential for $G$, as well as the existence of baryon
and lepton number violating interactions of dimensions 4 and 5
(apart from the dimension 5 operator responsible for neutrino
masses). The strong interaction scale, $M$, of the ${\cal G}$ theory
had to be of order $1$ TeV in order to obey constraints on chargino
masses.

CSB was implemented by adding an R violating term
$$\int d^2 \theta \lambda^{1/4} M_P^2 f(G/M_P) ,$$ to the
Lagrangian, where $\lambda$ is the cosmological constant.  When $G$
is small, this leads to an effective potential
$$V \sim \lambda^{1/4} M_P (K^{G\bar{G}} (G/M , \bar{G}/M )
| f_1|^2 - 3 |f_0|^2),$$  where $f_0$ and $f_1$ are the first two
coefficients in the Taylor expansion of $f(x)$ around the origin.
$K^{G\bar{G}}$ is the inverse Kahler metric generated for $G$ by the
strong dynamics of the ${\cal G}$ theory.  If the potential has a
minimum at $G \sim M$ then SUSY is broken.  $f_0$ is fine tuned (its
magnitude is of order one) to make the low energy effective
cosmological constant equal to $\lambda$, which is a high energy
input.

These models had many attractive features.   They gave an acceptable
pattern of SUSY breaking and solved the SUSY flavor and CP problems.
They provided a $\mu$ term of the right order of magnitude.   $SU(2)
\times U(1)$ breaking was partly determined by strong interaction
physics and the Higgs mass bounds of the MSSM were violated.
Finally, if the ${\cal G}$ theory has automatic CP conservation,
then the strong CP problem is solved, without axions or massless up
quarks.

The only problem with this class of theories is that no one has
found any members of the class.   In order to keep $G$ massless in
the $\lambda = 0$ limit, the ${\cal G}$ theory had to preserve R
symmetry (no spontaneous breaking) and was not allowed to have a low
energy field of R charge 2.  In all known examples, the $R = 2$
bilinear which couples to $G$ in the microscopic theory, appears as
an elementary field below $M$ and ruins the SUSY breaking mechanism.

In this paper we examine a model with a specific choice of the
${\cal G}$ gauge theory.  We then find that the addition of two
other singlets, $S,T$, with R-charge $2$, leads to a state with SUSY
breaking. The low energy, non-gravitational, effective field theory,
also contains SUSic minima.  The rules of CSB tell us to add a
constant term to the superpotential to tune the value of the
potential at the SUSY violating minimum to $\lambda$.  Given this
tuning, SUSic minima have negative energy density of order
$\lambda^{1/2} M_P^2 $.   We argue that transitions between the dS
space and negative energy Big Crunches are so improbable that it is
unlikely that any observer can survive long enough to experience
them.   The probability that an observer will experience a Big
Crunch before it is destroyed in some other fashion is of order
$e^{- c (RM_P)^2 }$, where $R$ is the radius of our dS horizon and
$c$ is a constant of order one.

In the next section we introduce an explicit and fairly unique
candidate model, the Pentagon, for the low energy sector of the
theory of Cosmological SUSY Breaking.   We then explore some of its
properties.  In section 7 we show that by a simple change of quantum
numbers we can convert the model into a model of dynamical breaking
of SUSY and electroweak gauge symmetry.  This section was motivated
by a remark of M. Dine. The two interpretations of the model are
phenomenologically similar, and differ mainly in their explanation
of the absence of certain terms in the effective action.  There is
however a key difference with regard to a pseudo-Nambu-Goldstone
boson, which plays the role of a QCD axions, and which is an
inevitable consequence of this model. In the DSB interpretation of
the model this PNGB gets mass only from the QCD anomaly.  It is a
low scale QCD axion, and is ruled out by experiment.   It may be
possible to solve this problem in the CSB interpretation of the
model by breaking the symmetry through a dimension $5$ operator.
This can raise the mass of the axion, and might explain its absence
in experiments done so far, and its lack of an effect on stellar
evolution.  At the moment, it appears that the requisite change in
mass can be achieved only by setting the scale of the Pentagon gauge
interactions at about $3$ TeV, which seems to require a fine tuning
of order one percent in dimensionless couplings in order to be
consistent with the scale of electroweak symmetry breaking.  This
PNGB is either one of the worst phenomenological vulnerabilities of
the Pentagon model, or a promising experimental signature.

 The dual DSB/CSB
interpretations point up the key new features of the model.  The
first is a new mechanism for electroweak symmetry breaking.   The
second is the use of {\it meta-stable} SUSY violating minima in a
field theory which has super-symmetric solutions in the absence of
gravity.   The c.c. is chosen to be very small and positive at the
meta-stable minimum, and this renders tunneling amplitudes to the
negative energy density region of the potential unobservably small.

Some brief comments about experimental signatures are in the
conclusions.   The paper also contains two appendices. The first is
devoted to an anthropic discussion of various coincidences in the
physics and cosmology of this model. The second contains some
calculations relevant to the structure of electroweak symmetry
breaking.

I end this introduction with a caution to the reader.   At a number
of points in the exposition I will have to assume that certain
couplings vanish.  In assessing the plausibility of these
assumptions, it is important to keep in mind the logic underlying
the present paper.  I am {\it assuming} that a quantum theory of dS
space exists, compatible with the hypothesis of CSB.  This paper is
an attempt to construct a low energy model compatible with that
assumption.  Thus, if it is necessary to assume certain couplings
vanish in order to find a SUSY violating minimum, then this will
follow automatically from the rules of the underlying theory. We do
not have to have symmetry explanations of every vanishing coupling.
Nonetheless, like a good little effective field theorist, I have
tried to find symmetry explanations.  The reader will judge how well
I succeeded.

Somewhat more disturbing are couplings that are required to be small
only for phenomenological reasons.   I have found symmetries which
forbid all of these at vanishing cosmological constant. The most
important of these is an R symmetry, which {\it must} be broken when
the cosmological constant is non-vanishing. One must then worry
about whether the dangerous couplings reappear.   The mechanism of R
symmetry breaking involves the still mysterious degrees of freedom
on the horizon of dS space\cite{susyhor}.   Thus, the answer to the
question about phenomenologically dangerous R violating couplings
(which don't disturb the SUSY breaking minimum) is not one which can
be answered in low energy effective field theory.

\section{\bf The Beast Whose Number is $555$}

In order to generate gaugino masses consistent with experimental
bounds, the ${\cal G}$ theory must contain chiral fields charged
under $SU(1,2,3)$.  To preserve coupling unification, it is best to
add full $SU(5)$ multiplets, so $SU(5)$ must be an anomaly free
subgroup of the flavor group of the theory.   We will choose the
$SU(5)$ supersymmetric gauge theory with $5$ flavors, and will call
this new gauge theory the Pentagon\footnote{Quintessence is already
in use, and Pentagram would be dangerous in the current funding
climate.} and the matter fields $P_i$ and $\tilde{P}^i$ the {\it
pentaquarks}\footnote{Although this term is already used for another
concept, it is my impression that we won't have to worry about
confusion much longer.} or p-quarks for short (The indices on
pentaquark fields describe their transformation under the
$SU(5)\times SU(5)$ flavor symmetry of the Pentagon model, and we
hide indices referring to the strong Pentagon gauge group).

We will constantly use the seminal results of Seiberg\cite{ns} on
the non-perturbative structure of SUSY QCD. In SUSY QCD with $N_F =
N_C$ the quark and anti-quark superfields have R charge zero under
the anomaly free $U(1)$ R symmetry.   We choose the action on the
pentaquarks of the exact $Z_N$ R symmetry required by the rules of
CSB, to be a subgroup of the anomaly free $U(1)_R$. That is, we
assume that the full theory of the universe, in the limit $\lambda
\rightarrow 0$, has an exact discrete symmetry, which acts on the
fields of this model as a discrete subgroup of the anomaly free $R$
symmetry.  In order to conform to the literature, we will call the
dynamical scale of the Pentagon $\Lambda_5$ rather than $M$

When the standard model couplings are turned off, the effective
theory below $\Lambda_5$ has a chiral superfield ${\cal M}_i^j = P_i
\tilde{P}^j$ in the $(5,\bar{5}) $ of $SU(5)_L \times
SU(5)_R$\cite{ns}. All components of this matrix have R charge 0.
Under the diagonal $SU(5)$ it breaks up into the direct sum of a
singlet and an adjoint. It is convenient to work with the two
standard model singlets ${\cal M}_t = {\rm tr}\  I_t {\cal M}$  with
$t = 2,3$. $I_t$ are the two orthogonal $SU(1,2,3)$ invariant
projectors in the $5$ representation of $SU_V (5)$.

There are two more composite standard model singlets, which are
massless when $\lambda = 0$.  These are the P-baryon, $B$, and
anti-P- baryon, $\bar{B}$. They have P-baryon number $\pm 5$.
P-baryon number is an exact symmetry of the low energy theory, which
we presume broken via irrelevant terms of order $1/M_U$ or $1/M_P$.
These are our dark matter candidates\cite{bmo}, and we will see that
they obtain mass once $\lambda$ is turned on.

Finally, we will introduce two chiral superfields, $S$ and $T$,
singlet under all gauge groups, and elementary at scales above
$\Lambda_5$. They have $R$ charge $2$, and Yukawa couplings

$$ \int d^2\theta\ \{ S [  g_S ( {\cal M}_2 + {\cal M}_3) \Lambda_5 + g_{\mu} H_u H_d] $$
$$+ T[ g_T (-{3\over 2}{\cal M}_2 + {\cal M}_3) \Lambda_5 ] \}.$$

If the R symmetry group is $Z_N$ with $N \geq 3$ and $N \neq 4$,
then R symmetry does not allow any renormalizable terms involving
only $S$ and $T$ in the Lagrangian. It would allow terms linear in
$S$ and $T$, multiplied by a function of $G$. However, there is a
discrete symmetry group, F, which forbids these. The Pentagon gauge
interactions leave a discrete subgroup of the axial $U_{AP} (1)$ of
the penta-quarks unbroken. By giving $S,T,H_u, H_d$ and the quarks
and leptons of the standard model charge under an appropriate
subgroup $F$ of this discrete group, we can make it a symmetry of
the whole Lagrangian. The terms involving $G$ break this symmetry,
and so must vanish when $\Lambda = 0$.

 Note also that, since we have
insisted that $H_u H_d$ have R charge 0, the action of the exact
discrete R symmetry on ordinary quarks, must combine a discrete
subgroup of the anomaly free $U_R (1)$ of SUSY QCD with six flavors,
with a subgroup of the discrete $Z_{12}$ subgroup of $U_A (1)$,
which is left unbroken by QCD instantons.

We have omitted a possible coupling of the form $T H_u H_d$, which
will turn out to be crucial to our discussion of CP violation.   One
possible excuse for this is that the couplings we have kept are
compatible with $S$ being a singlet, and $T$ a component of an
adjoint under the $SU_V (5)$ unified group, which is restored
(probably in higher dimensions) at the scale $M_U$.   Although
$H_{u,d}$ nominally come from a $[5]$ and $[\bar{5}]$, and can
couple to both a singlet and an adjoint, there are also ways to
prevent such a coupling at the unification scale.
Non-renormalization theorems will prevent it from being generated as
we scale down from $M_U$ to $\Lambda_5$.

On the subspace of moduli space where only ${\cal M}_t$ are
non-zero, the modified moduli space constraint yields
$${\cal M}_2^2 {\cal M}_3^3 = \Lambda_5^5 .$$  It is easy to see that there
is a supersymmetric vacuum with $SU(2)\times U(1)$ broken at the
scale $\Lambda_5$.  This follows from the moduli space constraint,
and $F_S = 0$.

The five flavor $SU(5)$ gauge theory (the Pentagon), its implicit
coupling to the SSM, and the Yukawa coupling of the singlets $S$ and
$T$ constitute the new features of our model.   The rest of the
model is just the SSM, without a $\mu$ term or soft SUSY breaking
terms, plus the Goldstino field $G$, which at the moment is a
completely decoupled massless chiral superfield. The full gauge
group of the model is $SU_P(5) \times SU_V(5)$, though only the
$SU(1,2,3)$ subgroup of $SU_V (5)$ is visible at low energies. The
full group is probably only realized in higher dimensions. The $P_i$
are in the $[5,\bar{5}]$ of this product group, and the
$\tilde{P}^i$ are in the $[\bar{5}, 5]$.  Quark/lepton fields are in
three copies of the $[1, \bar{5} + 10]$.  The Higgs fields are an
incomplete multiplet, perhaps arising from a $[1, 5 + \bar{5}]$. The
$S$ and $G$ fields are singlets of both groups and the $T$ field may
be a remnant of a $[1,24]$. This model is supposed to be a complete
description of low energy physics when $\lambda = 0$.

When we turn on $\lambda$, we add new $R$ and $F$ violating terms to
the effective action.  These terms are the {\it deus ex horizonae},
and we know very little about their nature, except that they scale
to zero with $\lambda$, and they must give rise to a gravitino mass
of order $\lambda^{1/4}$, and a c.c. of order $\lambda$. In
\cite{susyhor} I gave a hand waving prescription for calculating
these terms in terms of Feynman diagrams whose internal lines can
interact with the horizon degrees of freedom. Such diagrams mix up
infrared properties of the bulk, with Planck scale physics near the
horizon. Each of the R violating terms will scale with a
characteristic {\it critical exponent} $({{\lambda^{1/4}}\over
M_P})^p$ as the c.c. goes to zero. It will also have some dependence
on the IR scales of the theory, which will determine the number of
powers of the Planck mass in the coefficient.   Since the c.c. is so
small, it is plausible that only one R violating term dominates most
of the physics\footnote{I am not counting here the R violating
constant in the superpotential, whose role is to tune the c.c. to
$\lambda$.}.

For phenomenological reasons, and reasons of simplicity, I will make
the assumption that this unique R violating term has the form
$$\int d^2 \theta\ [g_G G \Lambda_5({\cal M}_2 + {\cal M}_3)  + \lambda^{1/4} M_P^2 f_0].$$
We will see that this leads to SUSY breaking of the order of
magnitude required by the hypothesis of CSB if $g_G \sim
{{\Lambda^{1/4} M_P} \over {\Lambda_5^2}}$.   The mechanism for SUSY
breaking in the effective field theory is similar to that of
Intriligator and Thomas\cite{it}. One of the inverse powers of
$\Lambda_5$ in this formula disappears when we go above the scale
$\Lambda_5$ and write the coupling as a Yukawa coupling to
penta-quarks. The other would be quite mysterious in conventional
effective field theory. It is plausible because of the IR dependence
of the diagrams of \cite{susyhor}. Note also that there is no way
that we can take $\Lambda_5 \rightarrow 0$. The ratio of $\Lambda_5$
to the Planck scale is completely fixed in the $\lambda = 0$
limit\footnote{This means that the phenomenological constraint
$\Lambda_5 \sim 1$ TeV, is an implicit challenge to the underlying
quantum gravity model. That model must predict the ratio of
$\Lambda_5$ to the Planck scale.}. $\lambda$ is the only tunable
parameter in the model.

It should be emphasized that we have gone beyond the rules of CSB at
this point.   We could have assumed that CSB was implemented by a
superpotential term\footnote{A term much less offensive to the
effective field theorist, because it contains no inverse powers of
the IR scale $\Lambda_5$.} $\lambda^{1/4} M_P G$. $G$ would be
stabilized at the unification scale by invariant terms of the form
$\delta W = {T^p \over {M_U^{p-3}}} f(G/M_U),$ with $f$ a bounded
function (see below). This would give rise to a hidden sector model
with the scale of SUSY breaking in the standard model far lower than
the electroweak scale.  Our decision to instead introduce SUSY
breaking via an order 1 Yukawa coupling, $g_G$, is based frankly on
phenomenology.   We will have to understand a lot more than we do
now about the underlying mechanism for CSB, before we can judge
whether the present model can be derived, rather than postulated.

\section{Baryon and lepton number}

A central element in CSB is the discrete R symmetry that guarantees
Poincare invariance in the the limiting model. This can be put to
other uses.   In \cite{susycosmophenoIII} I showed that it can
eliminate all unwanted dimension $4$ and $5$ baryon and lepton
number violating operators in the supersymmetric standard
model\footnote{Here we are invoking the hypothesis made above that R
violating terms other than $g_G$ scale with high powers of the c.c.
and are negligible.}. The interaction $\int d^2 \theta\ H_u^2 L^2$,
should not be forbidden by R.  We will adopt the philosophy of a
previous paper and insist that the texture of quark and lepton
Yukawa couplings, as well as neutrino masses, are determined by
physics at the unification scale.

We will choose the R charge of SSM fields to be independent of
quark and lepton flavor, and denote it by the name of the
corresponding field. All R charges are to be understood modulo
$N$, where $Z_N$ is the R symmetry group. Flavor dependent R
charges would require many important Yukawa couplings to vanish,
and the corrections to the R symmetric limit are too small to
account for the non-zero values of these couplings.

The condition that the standard Yukawa couplings are allowed by R
symmetry is

\eqn{yukallow}{L + H_d + \bar{E} = Q + H_d + \bar{D} = Q + H_u +
\bar{U} = 2.} As noted above, the Yukawa couplings of $S$ and $T$
require \eqn{sallow}{H_u + H_d = 0.} Note that this condition
forbids the standard $\mu$ term $\int d^2 \theta\ H_u H_d$. We will
also impose $2L + 2H_u = 2$ to allow the dimension $5$
superpotential responsbile for neutrino masses. The renormalizable
dynamics of the Pentagon gauge theory preserves all flavor
symmetries of the standard model. This forbids the generation of the
neutrino mass superpotential with coefficient ${1 \over \Lambda_5}$.
As emphasized in \cite{susycosmophenoIII}, we imagine the neutrino
mass superpotential, and the texture of the quark and lepton mass
matrices, to be determined by physics at the scale $M_U$, probably
via a Froggat Nielsen mechanism.

Dimension $4$ baryon and lepton number violating operators in the
superpotential will be forbidden in the limiting model by the
inequalities

\eqn{nobla}{2 L + \bar{E} \neq 2} \eqn{noblb}{2 \bar{D} + \bar{U}
\neq 2, } \eqn{noble}{L + Q + \bar{E} \neq 2 .} Absence of
dimension $5$ baryon number violating operators requires
\eqn{noblc}{3Q + L \neq 2}\eqn{nobld}{3Q + H_d \neq
2}\eqn{noblf}{\bar{E} + 2 \bar{U} + \bar{D} \neq 2 ,}

The condition that there be no baryon number violating dimension
$5$ D-terms is that none of $ Q + \bar{U} - L ; $ or $U + E - D$,
vanishes.

If we solve for $L,\ \bar{D},\ \bar{U},\ \bar{E}$ in terms of $Q$
and $H_u$, the inequalities become \eqn{noblg}{3Q - H_U \neq
1,2,4,5}\eqn{noblh}{Q + H_U \neq  0 .} These are all understood
modulo $N$\footnote{Actually, we have not written the most general
solution.  In solving the equation for $\bar{E}$ we neglected a term
that vanishes modulo $N/2$ but not modulo $N$.}.  If $N = 3$, a
possible solution is $H_u = 0$, $Q = 1,2$.   If $N \geq 4$, $H_u =
0$, $Q =1$ is always a solution, and there are others. Thus, like
the models of \cite{susycosmophenoIII}, the R symmetry of the
Pentagon model can be chosen to forbid all dangerous baryon and
lepton number violating terms.

\section{CP Violation and Flavor}

Except for the standard Yukawa couplings, the entire low energy
effective Lagrangian is invariant under the $SU(3)_Q \times
SU(3)_{\bar{U}} \times SU(3)_{\bar{D}}$ flavor group of the standard
model.   Thus, if we imagine that all higher energy physics comes in
above the unification scale, then the Pentagon model has a GIM
mechanism and predicts flavor changing neutral currents within
experimental limits.   Therefore, we adopt the philosophy of
\cite{susycosmophenoIII}, according to which the origins of flavor
and neutrino mass physics reside at the unification scale.

We begin our discussion of CP violation in the effective theory
above the scale $\Lambda_5$.  We will continue to make the
assumption that there is only one significant $R$ violating
coupling, $g_G$, in the low energy Lagrangian at this scale.   This
coupling can be made real by a phase rotation of the Goldstino
superfield $G$.   The problem of CP violation can then be addressed
in the $\lambda = 0$ limit. The Lagrangian is that of an
$SU(5)\times SU(1,2,3)$ gauge theory, coupled to two singlet chiral
fields $S,T$ with superpotential
$$W = g_S S  [P_i \tilde{P}^i] + g_{\mu} S H_u H_d + g_T T [Y^i_j
P_i \tilde{P}^j ]$$
$$ + \lambda_U^{mn} H_u Q_m \bar{U}_n +
\lambda_D^{mn} H_d Q_m \bar{D}_n + \lambda_L^{mn} H_d L_m \bar{E}_n)
.$$  The $SU(5)$ and $SU(1,2,3)$ gauge indices are implicit in these
formulae.  The matrix $Y$ is defined by $Y \equiv I_3 - {3\over 2}
I_2$. There is also a dimension five term which will generate
neutrino masses when $H_u$ gets a VEV, and a variety of other terms
scaled by inverse powers of $M_U$ or $M_P$. We will not discuss
questions of flavor and CP violation in the lepton sector in this
paper.

We begin our discussion of CP violation by performing $U_A (1)$ and
$U_{AP} (1)$ transformations to eliminate the imaginary parts of the
two strong gauge couplings $\theta_3$ and $\theta_5$.   From this
point on we will only perform anomaly free transformations on chiral
superfields, so these angles will remain zero.

We have noted that the only terms in the effective Lagrangian which
violate the $SU(3) \times SU(3) \times U_B (1)$ flavor symmetry of
the standard model, are the Yukawa couplings $\lambda_U$ and
$\lambda_D$. We can use the flavor group to eliminate all of the
phases in $\lambda_{D,U}$ except the usual CKM angle and a single
overall phase.

Now we perform the following sequence of transformations:

\begin{itemize}

\item An anomaly free $U_R (1)$ transformation, with angle $\alpha$.

\item A rotation of $S$ by phase angle $\beta$.

\item A rotation of $T$ by phase angle $\gamma$.

\item  A common phase rotation of $H_u$ and $H_d$

\end{itemize}
It is easy to see that these rotations can be chosen to eliminate
all phases in $g_{T,S,\mu}$ and ${\rm arg\ det}\ [ \lambda_u
\lambda_d ]$.    The effective potential, which includes effects of
the CP conserving Pentagon gauge interactions, is CP invariant.  We
will assume (this must be checked dynamically) that the minimum we
find after SUSY breaking does not violate CP spontaneously.   {\it
If that is the case, the Pentagon model solves all CP problems of
SUSY, as well as the strong CP problem.}   The only CP violating
phase in the renormalizable effective theory at scale $\Lambda_5
\sim 1$ TeV is the usual phase in the CKM matrix. However, we will
see below that in effect what we have constructed is a
supersymmetric version of the Peccei-Quinn axion model. The model
has a pseudo-Nambu-Goldstone boson whose mass (taking into account
only the renormalizable couplings in the Lagrangian) comes
predominantly from QCD.  This is a Peccei-Quinn-Weinberg-Wilczek
axion, and is ruled out by a combination of laboratory experiments
and stellar physics.

We can solve the laboratory problems by taking $\Lambda_5 \sim 3
TeV$ (which may also alleviate possible problems with precision
electroweak data, but may require fine tuning to get the electroweak
scale right).   The stellar problems can be resolved by raising the
axion mass with a dimension $5$ operator.   Unfortunately, the phase
of this new coupling, probably reintroduces the strong CP problem.

\section{$SU(2) \times U(1)$ and SUSY breaking}

We have noted above that $SU(2) \times U(1)$ is already broken when
$\lambda =0$.   The nominal breaking scale is of order $\Lambda_5$,
and we see that $\Lambda_5$ is required to be $1$ TeV or smaller.
However, the uncertainties introduced by the strong Pentagon
interactions, as well as by the new dimensionless couplings we have
introduced, will prevent us from getting more than order of
magnitude estimates of the masses of various particles. The
electroweak scale is related to $\Lambda_5$ by a function that
depends on the couplings $g_S, g_{\mu}$, and the non-perturbative
dynamics, which gives an expectation value to ${\cal M}_t$.  We will
see below that there are phenomenological reasons to want $\Lambda_5
\sim 3$ TeV , in order to avoid experimental problems with an
axion-like particle.

We demonstrate in Appendix 2, that once SUSY breaking is taken into
account, there is a range of parameters for which $SU(2) \times
U(1)$ is broken, and every physical component of the Higgs fields,
$H_{u,d}$ gets a mass of order $\Lambda_5$, independent of the
standard model gauge couplings. Thus, the conventional SUSY upper
bounds on the Higgs mass do not apply to this model. Precision
electroweak fits put an upper bound of about $200$ GeV on the Higgs
mass (though this is in the absence of new physics, which is present
in abundance in the Pentagon model).

A possible problem with a scale as low as $1$ TeV is the detailed
agreement of precision electroweak fits with the unadorned standard
model. At the moment, I am too calculationally challenged to
determine whether this is a problem for the Pentagon model, but
increasing the scale to $3$ TeV will help. On the positive side, it
is worth pointing out that the pattern of electroweak symmetry
breaking picked out by the Pentagon is precisely the same as that
conventionally attributed to the effect of the standard model D
terms.

In making the above analysis, I have used the fact that low energy
physics in the Pentagon model is exactly CP invariant (up to the CKM
angle), but have made the additional assumption that the potential
chooses a CP conserving minimum for the fields.   The latter
assumption is certainly plausible, but needs further investigation.

\subsection{SUSY breaking}

Recall that the c.c. is a tunable parameter.   Our best guess at the
dynamical structure of this model is obtained in the limit that
$\lambda$ is much smaller than its observed value.   In this regime,
we can continue with the effective field theory analysis, which
reveals the full qualitative structure.   We emphasize that this
analysis breaks down for the observed value of $\lambda$, where we
must deal with the full dynamical complication of the strongly
coupled Pentagon model. However, the small $\lambda $ analysis gives
a very attractive phenomenological picture, and we can hope that it
survives a more rigorous treatment.

In the limit of small $\lambda$ we deal with an effective field
theory for the moduli, below the scale $\Lambda_5$.   This theory is
determined by a Kahler potential, and a superpotential
$$W = S[g_S ({\cal M}_2 + {\cal M}_3)\Lambda_5 + g_{\mu} H_u H_d] + g_T T ({3\over 2}{\cal M}_2 - {\cal M}_3)\Lambda_5$$
$$ + g_G G ({\cal M}_2 + {\cal M}_3)\Lambda_5 + H_u Q^{T} \lambda_u \bar U + $$
$$H_d (Q^{T} \lambda_d \bar {D} + L^{T} \lambda_l \bar{E}$$   There is also a constraint on the moduli space
$${\rm det} {\cal M} - B \bar{B} = \Lambda_5^5 .$$

Apart from the $SU(1,2,3)$ gauge group of the standard model, only
discrete subgroups of the $R$ and flavor groups are exact symmetries
of the world (when $\lambda = 0$), but we are neglecting terms
inversely proportional to the unification scale. For constraining
the dynamics of the strongly coupled Pentagon theory, we can
consider its flavor group to be a continuous symmetry group.

It is convenient to write the first term in the superpotential as
${\rm tr}\  (N {\cal M})$, where the spurion field $N$ transforms in
the $(\bar{5}, 5)$ representation of the flavor group.  Later, we
will set $N = (g_G G + g_S) I  + g_T T Y$, where $Y$ is weak
hypercharge. The most general invariants we can make are ${\rm tr} [
(N {\cal M})^k , ({\cal M}{\cal M}^{\dagger})^k , (N N^{\dagger})^k,
({\cal M}^{\dagger} N^{\dagger} )^k ]$, with $ 1 \leq k \leq 5$.

We will restrict attention to the submanifold of moduli space where
${\cal M} = {\cal M}_3 I_3 + {\cal M}_2 I_2$. We will verify later
that the excitations of other moduli normal to this submanifold, are
massive\footnote{Though, to be honest, we will not verify the
absence of tachyons.}. The same will be true for the baryonic
components of the moduli space. On this submanifold, there are more
invariants than variables, and the flavor symmetry of the Pentagon
does not restrict the form of the Kahler potential.  It is
constrained only by CP invariance.

The $F$ terms of the four independent chiral superfields,
$S,T,G,M_3$ are

$$F_S =  g_S \Lambda_5({\cal M}_2 + {\cal M}_3 ) + g_{\mu} H_u H_d,$$
$$F_{T} = g_T \Lambda_5 ({\cal M}_3 - {3\over 2} {\cal M}_2) ,$$
and
$$F_G =  g_G ({\cal M}_2 + {\cal M}_3) \Lambda_5 ,$$
$$F_3 = (g_S \Lambda_5 S + g_G \Lambda_5 G) (1 + {\cal M}_2^{\prime} ) + g_T T \Lambda_5 ( 1 -
{3\over 2} {\cal M}_2^{\prime}) .$$ In these equations, ${\cal M}_2
= \Lambda_5 ({\Lambda_5 \over {{\cal M}_3}})^{3/2},$  and ${\cal
M}_2^{\prime}$ is its derivative with respect to ${\cal M}_3$.  The
$F$ terms of $H_{u,d}$ are $$F_{u,d} = g_{\mu} S H_{d,u}$$ (we set
the squark and slepton VEVs to zero). It is clear that $F_S = F_T =
F_G = 0$ is an inconsistent set of equations when $g_G \neq 0$, so
SUSY is broken. When $g_G$ is small, the energy will be minimized
with all $F$ terms $\leq F_G$. For the actual value of $\lambda$,
this leads to energies of order $\Lambda_5$ or even greater (by a
small factor).  Thus, effective field theory is not a good tool in
the regime of phenomenological interest. Nonetheless, since it is
all we have available, we can hope that it gives the right
qualitative physics.

In the low energy approximation in which we are working, the
supergravity formula for the potential reduces to

$$V = \mu^4 (K^{i\bar{j}} F_i \bar{F}_{\bar{j}} - |f_0|^2),$$
where $\mu^2 \sim \lambda^{1/4} M_P $.   The rescaled chiral fields
$X^i$ which appear in this formula are $g_G G/\Lambda_5 , g_G^2 (S -
S^0)/\Lambda_5, g_G^2 (T - T^0)/\Lambda_5, g_G^2 ({\cal M}_3  -
{\cal M}_3^0 )/\Lambda_5$, and $g_G^2 (H_{u,d} - H_{u,d}^0
)/\Lambda_5$. The symbols with zero superscripts refer to the value
of the corresponding field at the supersymmetric minimum, when
$\lambda = 0$.   The linear, rather than quadratic dependence on
$g_G$ in the $G$ dependence of the potential, is due to the fact
that this field has a flat potential when $g_G = 0$. To leading
order in electroweak gauge couplings, the Kahler metric depends on
the Higgs fields only through the combination $g_{\mu} S H_u H_d $.
When $g_G$ is small, $S$ is of order $g_G^2$ and the overall scale
of the Higgs VEVS is fixed by the standard model D-terms.   Some
components of the Higgs will be lighter than $\Lambda_5$ by factors
of standard model couplings. However, for $g_G$ of order its
experimental value, all physical Higgs fields have mass order
$\Lambda_5$, independent of the standard model couplings. We discuss
this more extensively in Appendix 2.

For a range of parameters, we will have both $SU(2) \times U(1)$ and
SUSY broken, with the electroweak scale of order $\Lambda_5$, and
the SUSY breaking scale of order $\sqrt{g_G} \Lambda_5 $. The ratio
of the two scales is of order one for the observed value of the
c.c.\footnote{We discuss this coincidence in Appendix 1.}.  The
potential for the Higgs fields depends on their ratio, ${\rm tan}
\beta \equiv {|H_u| \over |H_d|}$, only through ${\rm sin} 2 \beta$
and is minimized when the ratio is one, throughout the parameter
range in which electroweak symmetry is broken. Thus, up to radiative
corrections, the model predicts ${\rm tan} \beta = 1$.

One worrisome feature of the model is that, as $g_G \rightarrow 0$,
the minimum of the potential for $G$ wanders out to infinity.  We
may worry that SUSY will be exactly restored for large values of
$G$, which would make the low energy model incompatible with the
basic principle of CSB.  In fact, this does not happen.   The
discrete symmetry which forbids renormalizable couplings of $T$ and
$G$, will allow terms of the form $W = {T^p \over M_U^{p-3}} f(G/
M_U) $.   When $G$ is of order $M_U$ the function $f$ will be order
$1$.   Recalling that $T \sim M$ we find an approximate form for the
potential when $g_G \ll 1$ and $G \sim M_U$.
$$V \sim  \lambda^{1/2} M_P^2 |(g_G G)|^2 + {\rm Re}[\lambda^{1/4} M_P \Lambda_5^2
f^{\prime} (G/M_U)].$$   If the second term dominates the first we
find a minimum at $f^{\prime} (G_0 / M_U) = 0$, with a SUSY breaking
$F$ term of order $\lambda^{1/4} M_P$.   The condition for this to
happen is
$$\lambda^{1/2} M_P^2 < \Lambda_5^4 ({\Lambda_5 \over M_U})^p , $$ which will always
be satisfied for small enough $\lambda$.   That is, the system will
stabilize, with the right scale of SUSY breaking, with a unification
scale VEV for $G$, in the limit of asymptotically small $\lambda$.

\subsection{Massive moduli and light superpartners}

When $\lambda = 0$, the Pentagon model has a plethora of exactly
massless moduli.   These all get mass for nonzero c.c.   To estimate
the magnitude of these masses, consider quartic terms in the Kahler
potential
$$\delta K \sim {{a G\bar{G} \bar{X}X }\over M^2}.$$  The scalar
components of $X$ will get mass of order
$$m_X \sim {{F_G} \over \Lambda_5^2}.$$  Terms in $K$ of the form
$$b \bar{G} D_{\alpha} X D^{\alpha} X  + h.c. ,$$ give similar masses to
the fermionic components.   For the observed value of $\lambda$, and
$\Lambda_5 \sim 1$ TeV, these are all in the TeV range.   The
observation of these P-hadrons, with the standard model quantum
numbers attributed to them by the model, would be a spectacular
experimental signature. Among these particles are the components of
the $P-baryon$.   These are standard model singlets and make an
excellent dark matter candidate\cite{bmo}, since the P-baryon number
is conserved up to interactions of order ${1\over {M_U^q}}$ with $q
\geq 1$. Although P-baryons cannot be thermal relics, a primordial
asymmetry in this quantum number could account for the observed dark
matter density. If the asymmetry were generated by the same physical
mechanism as the baryon asymmetry, we might even hope to find an
explanation of the dark matter to baryon ratio.

It is worth noting that pseudo-Nambu-Goldstone bosons associated
with spontaneous breaking of the axial $SU(5)$ generators by the VEV
of ${\cal M}_3$ are not in fact light.  $SU_A (5)$ is explicitly
broken by the combination of couplings $g_S , g_T$ even in the SUSic
limit, so these PNGBs will only be light if the Yukawa couplings are
small. There are a variety of phenomenological reasons to assume
that this is not the case.   One would need a full understanding of
UV boundary conditions at $M_U$ and of the renormalization group
equations in order to decide whether it is plausible that all of
these couplings are relatively large.   Note that they all involve
fields which are coupled to the asymptotically free Pentagon gauge
theory, which at least pushes things in the right direction.  In the
next subsection we will describe a pseudo-Nambu-Goldstone boson,
which follows from an approximate continuous R symmetry of our low
energy Lagrangian.

I have not been able to find an argument that all of the SUSY
violating scalar masses squared are positive.   This is another
dynamical assumption (local stability of the SUSY violating minimum)
for whose justification one must look to the solution of the
strongly coupled Pentagon theory\footnote{N. Seiberg suggested that
one check stability in an extreme region of moduli space, where the
Kahler potential is calculable.  Preliminary investigation suggests
that all squared masses are positive in this region, but the mass
matrix in this region is not related in any simple way to that in
the phenomenologically interesting region of moduli space.}.

The fact that when $\lambda$ is much smaller than its observed
value, the model predicts a lot of light neutral particles with
couplings to the standard model of order $1/\Lambda_5$, as well as a
lot of light charged particles, may be of interest for tightening
the anthropic {\it lower bound} on $\lambda$.  The change in the
properties of dark matter\footnote{The primordial asymmetry, and
fluctuation spectrum are determined by high energy physics
independent of $\lambda$.} with $\lambda$ will alter the nature of
galaxies, which for sufficiently small $\lambda$ will be purely
baryonic.  Light weakly coupled particles will alter the nature of
stars, and new light charged states (in addition to the squarks,
sleptons, and charginos) will change the nature of chemistry. We
will discuss this briefly in Appendix 1.

It appears that, apart from the axion,  all the states lighter than
$M$, will be standard model particles, and their superpartners.
Higgsinos, winos and zinos will have SUSic masses of order $g_{1,2}
\Lambda_5$, and will also get SUSY violating masses of order
${{\alpha_{1,2}}\over \pi} {\lambda^{1/4} M_P \over \Lambda_5} $.
Gluinos and photinos will get SUSY violating masses given by
analogous formulae, with appropriate standard model fine structure
constants. This suggests that some charginos may not be far above
their experimental lower bounds, and in the discovery range of LHC.
Similarly, sleptons and squarks will have masses related to
corresponding gaugino masses as they are in gauge mediated models.
The suggestion that sleptons are relatively light is exciting,
because in a low energy SUSY breaking model, like the Pentagon, they
have spectacular decays.

\subsection{A pseudo Goldstone boson}

The following $U_E(1)$ R transformation is a symmetry of the
classical Lagrangian of our model:  let $S,T,G$ have E charge $2$
and the penta-squarks E charge $0$.  We have to assign the E charge
$E_u = - E_d$ to the Higgs fields.  Invariance of the usual quark
Yukawa couplings requires that
$$E_Q + E_{\bar{U}} + E_u = 2 = E_Q + E_{\bar{D}} - E_u ,$$ whence
$$2E_Q + E_{\bar{U}} + E_{\bar{D}} = 4 .$$  The latter quantity is
all that we need to evaluate the QCD anomaly of the E symmetry,
which is non-vanishing.   $U_E (1)$ acts on the penta-quarks like
the anomaly free R symmetry and so has no Pentagon anomaly.

As a consequence, at the renormalizable level, the dominant source
of E breaking is the QCD anomaly.  We have also argued that the
fields $S,T$ get VEVs of order $\Lambda_5$, so there is a
pseudo-Goldstone boson, the axion,  with decay constant of order $1$
TeV, which gets mass from the QCD anomaly.  It is well known that
such a particle is problematic, both from the point of view of
terrestrial experiments, and of stellar physics.

We can try to solve this problem by breaking the E symmetry
explicitly using irrelevant operators from the unification scale. An
operator of dimension $d$ will give an axion mass of order
$$m_{ax}^2 \sim {{<{\cal O}_d>}\over {\Lambda_5^2 M_U^{d-4}}} ,$$ which
should be larger than the QCD induced mass

$$m_{axQCD}^2 \sim {{\Lambda_{QCD}^4} \over \Lambda_5^2}.$$  With $\Lambda_5 \sim 1$
TeV, and $M_U \sim 10^{16} $ GeV, this can only be satisfied if $ d=
5$ and the VEV is of order $\Lambda_5^d$.   In that case the mass is
about $10$ times the value expected from QCD.   There are no
dimension $5$ superpotentials constructed from the Pentagon degrees
of freedom, $S,T,$ and $G$, which violate the $E$ symmetry, but
preserve a discrete $R$ symmetry when $g_G = 0$.   However, if we
assume the fundamental discrete R charge of $G$ is zero, then we can
write a dimension $5$ Kahler potential term
$$\int d^4 \theta (P \bar{P} {G/M_U} h_a + c.c.) .$$  In section 7 we will
introduce an alternative interpretation of the low energy
Lagrangian, in terms of Dynamical SUSY Breaking (DSB). In this
version of the model the exact discrete R charge of $G$ is $2$, and
the dimension 5 term would not be allowed.  It would then appear
that the model has a low scale QCD axion, and is ruled out
experimentally. Thus, experiment mildly prefers the cosmological
interpretation of the origin of $g_G$\footnote{One could imagine, as
originally suggested by Dine, replacing the mysterious cosmological
breaking of R symmetry by conventional dynamical breaking.  However,
this always seems to produce $g_G \ll 1$, which is not
phenomenologically viable.}.

Of course, it is by no means certain that this irrelevant operator
solves the problem.   We are predicting a $100$ keV particle with
couplings to the standard model which are probably close to weak
interaction strength.  The QCD anomaly in $U_E (1)$ symmetry gives
the axion a mixing with the $\pi^0$ of order
$c{{F_{\pi}}\over\Lambda_5}$, with $c$ a number of order 1.   This
is enough to produce couplings to hadrons and photons which make the
axion a problem both for terrestrial experiments and stellar
evolution.

We can resolve the accelerator and reactor problems by postulating
that $\Lambda_5 / c > 3$ TeV.  The stellar problems can be resolved
by lowering the mass parameter in the dimension five operator by a
factor of $10$ (so that it has the same size as the parameter that
appears in the dimension five operator, which generates neutrino
masses), and taking $\Lambda_5 \sim 3 TeV$.   This gives an axion
mass $\sim 5 MeV$, which means that axions will not be produced in
most normal stars.   On the other hand, they couple strongly enough
to be trapped inside of supernovae.

This relatively high scale for the Pentagon gauge interactions is
probably also good for protecting the agreement of our model with
precision electroweak data.    On the other hand (see Appendix 2) it
seems to require fine tuning of dimensionless couplings with about
one percent accuracy, in order to get the correct value for the
electroweak scale.  It is clear that even if we succeed in curing
the problem of not seeing the axion in existing experiments, it
should be discoverable by an extension of axion searches.  It is
perhaps one of the most definite experimental signatures of our
model.

Finally, we note that the phase of the coupling $h_a$ will determine
the VEV of the axion field, and appears to reintroduce the strong CP
problem.   To be certain that this is the case, we should integrate
out physics at the scale $\Lambda_5$ with care, and directly compute
the neutron electric dipole moment in terms of ${\rm arg}\ h_a$, but
the result doesn't look promising.

\subsection{Coupling constant unification}

{}From the point of view of the standard model, the Pentagon gauge
theory adds matter in the $[5]$ of the standard Grand Unified
$SU(5)$.  This means that, at one loop, coupling constant
unification works the same way as it does in the Supersymmetric
standard model, with the same unification scale of about $2 \times
10^{16}$ GeV.  This is the scale $M_U$ that we have mentioned
repeatedly in the text.   The value of the unified couplings is
different, and since there are $5$ extra triplet-anti-triplet pairs
of chiral superfields in the Pentagon, the unified coupling is on
the edge of the weak coupling regime at the unification scale.
Indeed, we need a full two loop calculation, taking into account the
effect of the Yukawa couplings and threshold effects associated with
the strongly interacting Pentagon theory at the TeV scale in order
to assess whether perturbative unification truly occurs in this
model.  Note that the relatively strong unified coupling might bring
dimension 6 proton decay amplitudes into experimental range.

Note that it is this property of coupling unification which
specifies the Pentagon theory among all possible $N_F = N_C$ models,
as the likely candidate for a description of realistic particle
physics in the framework of CSB. For $N_C < 5$ we cannot embed the
standard model in a vector-like subgroup of the flavor group, while
for $N_C > 5$ we lose perturbative unification.

A significant feature of the Pentagon model is that QCD is not
asymptotically free above the TeV scale.   In fact, the scale at
which asymptotic freedom is lost scales like ${{\lambda^{1/4}
M_P}\over \Lambda_5}$. This is the TeV scale for the observed value
of $\lambda$.  Thus, the Pentagon model is a realization of old
ideas for explaining the relative proximity (on a logarithmic scale)
of the QCD and electro-weak scales\cite{parisi}.

\section{Vacuum decay}

We have explored a particular region of moduli space, and found a
SUSY violating minimum of the potential.  It may be meta-stable, and
that is of course crucial to our enterprise.   However, it is almost
certain that there are SUSic points when other moduli are turned on,
so the dS space we find is likely to suffer from Coleman-DeLucia
vacuum decay.   This decay is of no phenomenological significance.
Part of our prescription for finding a model, was to tune the  c.c.
at the SUSY violating minimuum to $\lambda $.    As a consequence,
the energy density at SUSic minima of the potential will be of order
$ - \lambda^{1/2} M_P^2 $ and there is likely to be a CDL instanton
describing the ``decay" of de Sitter space into a Big Crunch
universe.  A non-gravitational estimate of the action of this
instanton is $\sim {8\pi^2 \over g^2} $, with $g$ a small coupling,
and it could easily predict a lifetime longer than the age of the
universe. In fact, in \cite{etinf}I will argue that the
gravitational corrections to this instanton are large and that its
action is actually of order $I_{inst} \sim c {{M_P^4 \over
\lambda}}$. If this is correct, then the lifetime is one of those
times so long that the unit one measures it in is irrelevant. It is
virtually the same number measured in current ages of the universe
as it is in Planck times. It is much more probable, by factors of
order $e^{c R^2}$\footnote{$R$ is the radius of dS space.}, for an
observer in dS space to be destroyed by the nucleation of a black
hole at the observer's position, than to experience CDL vacuum
decay.

From a more fundamental point of view however, this decay is
troubling, because we thought we were finding an effective
description of a stable dS space.   In a future paper\cite{etinf} I
will argue that the CDL instanton does not really represent a decay.
Rather, it describes a highly improbable statistical fluctuation in
which the system temporarily goes into a very low entropy state
(like all the air in the room gathering in a corner).

\section{The Pentagon as a model of dynamical SUSY breaking}

In this section, I want to show how the Pentagon model can be turned
into a model of dynamical SUSY breaking, abandoning its connection
with de Sitter space and quantum gravity\footnote{The idea of
severing the connection between the Pentagon model and CSB was
suggested by M. Dine.}. I have been reluctant to do this, because of
a belief in the fundamental connection between SUSY breaking and the
structure of space-time.  However, the resulting model is in some
ways simpler, and has fewer mysterious assumptions, than its CSB
motivated progenitor.

The DSB model, begins with the supersymmetric Pentagon model, but
changes the R charge assignment of $G$ to $R_G = 2$, so that $g_G$
can be non-zero from the outset. We view $G$ and $S$ as originating
from $SU(5)$ singlets at the unification scale, while $T$ comes from
a $[24]$. Both $S$ and $G$ can couple to $P_i \tilde{P}^i$ as well
as to $H_u H_d$, but we take can take linear combinations and define
$S$ to be the combination that couples to $H_u H_d$.   We omit the
coupling of $T$ to $H_u H_d$ to avoid a CP violating phase. As in
the CSB version of the Pentagon, we believe that this can be
explained at the unification scale and propagated to low energies by
the non-renormalization theorem. Recall that inclusion of this
coupling does not affect the dynamics of electroweak or SUSY
breaking, but only the strong CP problem.

Renormalizable polynomial interactions between $T,S,$ and $G$ are
forbidden by a $Z_N$ R symmetry with $4 \neq N \geq 3$.   Terms of
the form $\mu^2 X$, with $X = S,T,U$, are forbidden by the discrete
$U_AP (1)\times U_A(1)$ symmetry that forbade terms of the form
$Sf(G)$ in the previous sections. At the classical level, there is a
supersymmetric vacuum where all fields have zero VEVs. Once we take
into account the dynamical scale $\Lambda_5$ of the Pentagon theory,
this vacuum is replaced by a SUSY violating one. The scale of SUSY
breaking is determined by the smallest of the couplings $g_G, g_T,
g_S$ (in the low energy EFT approximation discussed above, $g_G$ was
always assumed smallest), multiplied by $\Lambda_5$.  $SU(2) \times
U(1)$ is broken at a scale of order $ \Lambda_5$.

As in the CSB version, there are SUSic points in moduli space, in
particular, a point where only the ${\cal M}_t $ components of the
meson field, as well as the product $B\tilde{B}$ are non-zero.  It
is important to check that the Pentagon dynamics actually produces a
locally stable SUSY violating minimum.  If this is the case, and we
tune the constant in the superpotential to make the potential of
order $\lambda$ at the SUSY violating minimum, then non-perturbative
instabilities are {\it at least} phenomenologically irrelevant, and
may not be instabilities at all when quantum gravity is properly
understood\cite{etinf}.

The existence of this alternative interpretation of the Pentagon
model has implications for the CSB interpretation as well.  It
highlights the existence of an accidental R symmetry of the low
energy model under which $G$ has R charge $2$.  From the low energy
point of view, this is a $U(1)$ symmetry, broken to a discrete group
only by non-perturbative QCD effects and the dimension 5 operator we
introduced to raise the axion mass. When $g_G$ is small, we have
used an effective field theory approach to estimate various scales
of symmetry breaking and particle masses. In particular, the SUSY
violating part of gaugino masses, is estimated by calculating the
gauge coupling functions for the standard model gauge multiplets.
This calculation is non-perturbative in the Pentagon interaction
strength, and one loop in standard model couplings\footnote{We have
not presented the calculation in this paper, but using the magic of
holomorphy, it reduces to a one loop calculation in the Pentagon as
well.}.   The $U(1)$ R symmetry implies that the gauge coupling
functions can only depend on the moduli ${\cal M}_t$ and not on
$S,T,G$.   It is therefore important that the F term of ${\cal M}_3$
be non-vanishing.  When $g_G = 0$, $F_3$ can be made to vanish by
choosing the otherwise free expectation values of $S$ and $T$.  When
we turn on $g_G$, SUSY is broken.  The potential has a complicated
dependence on $S$, $T$ and $G$, through the Kahler potential.  There
is no apparent reason for $F_3$ to vanish or be anomalously small.

One apparent advantage of the DSB interpretation of the Pentagon is
that within the realm of effective field theory in flat space, it is
consistent to assume that the discrete $R$ symmetry is
exact\footnote{This argument is somewhat naive.   It seems clear
that a consistent theory of dS space can have no exact symmetries
for a local observer, because charge can flow out through the
horizon.}, and only broken spontaneously at the scale $\Lambda_5$.
This leads to sufficient suppression of dangerous $B$ and $L$
violating interactions.   However, this is probably not consistent
with the size of $f_0$, the constant in the superpotential which
fine tunes the c.c. to $\lambda$.   Thus, whichever way we look at
it, the issues of SUSY breaking and the tuning of the c.c. are
intimately related.  In addition, this exact discrete R symmetry,
implies the accidental continuous $U_E (1)$ symmetry, and forbids
the dimension $5$ Kahler potential term which could prevent the PNGB
of this symmetry from being an experimentally challenged QCD axion.
Thus, the conservative effective field theorists preference for the
DSB interpretation of the model, leads in unpromising directions.

There have been several previous attempts, \cite{hitoshi}\cite{luty}
to break electroweak symmetry using supersymmetric dynamics.  Our
model most resembles that of \cite{luty} but differs in detail.  In
particular, we work on a part of the Pentagon moduli space which
preserves $SU(2)\times U(1)$, so that the VEVs of the elementary
Higgs fields are the primary source of electroweak breaking.   The
authors of \cite{luty} also considered, but rejected, the
possibility of gauge mediated SUSY breaking for the gauginos and
sparticles of the SSM.   From their point of view we are insisting
on a larger Yukawa coupling $g_G$ than might be considered natural.
Its value is motivated by CSB. If it should turn out that this
coupling is unrealistically large, we could, from the CSB point of
view, add a term  $\lambda^{1/4} M_P G$ to the superpotential (this
would not fit well in the dynamical context of the present section)
to achieve the value of the gravitino mass required by CSB. This
would spoil our solution of the strong CP problem.  Another major
difference between the present paper and previous work is the way in
which coupling unification is achieved.

\section{\bf Conclusions}

The ideas of CSB, when combined with the existence of the standard
model, and of coupling unification, lead to a rather unique model
for TeV scale physics, the Pentagon model.  This model adds only a
few new parameters to low energy physics.   In the SUSic limit,
these are just $g_S$, $g_T$, $g_{\mu}$ and the scale $\Lambda_5$ at
which the Pentagon gauge interactions become strong.  SUSY violation
adds one new parameter $g_G$.  The model automatically conserves CP,
except for the usual CKM angle. In particular, the QCD vacuum angle
vanishes. The model has a discrete $R$ symmetry which forbids all
baryon and lepton number violating operators of dimensions $4$ and
$5$, apart from the operator that gives rise to neutrino masses.

SUSY is broken by an $R$ violating superpotential, $(\lambda^{1/4}
M_P^2 f_0 + g_G G P_i \tilde{P}^i)$, which is attributed to
interactions with the horizon of de Sitter space (a {\it deus ex
horizonae}). $g_G$ is given by the formula ${{\lambda^{1/4}
M_P}\over \Lambda_5^2}$, which seems designed to induce heart
attacks in dedicated practitioners of effective field theory.  $f_0$
must be tuned to guarantee that the cosmological constant in the low
energy effective Lagrangian agrees with its fundamental input value.
This is the usual tuning we do in effective field theory, but here
motivated by the insight that $\lambda$ is a high energy input which
cannot be renormalized by low energy field theory effects.  The
other bizarre feature of the model is the unusual power law
dependence of the superpotential on $\lambda$ and $\Lambda_5$. $f_0$
does not seem to have a dramatic effect on particle physics, at the
level of approximation in which we have worked.

Our analysis is done in a limit where the cosmological constant is
even smaller than its observed value.   The effective composite
field theory valid in that limit, predicts that SUSY is broken at a
scale $\sqrt{\lambda^{1/4} M_P}$, while $SU(2) \times U(1)$ is
broken at a scale $\Lambda_5$.  The electroweak breaking mechanism
is very different from that of the SSM, and conventional Higgs mass
bounds do not apply.  The ratio of Higgs VEVs, ${\rm tan} \beta$ is
predicted to be close to one. The flavor and CP problems of SUSY are
completely solved, as well as the strong CP problem. In this model,
the origin of flavor and of neutrino masses has to do with physics
at very high energy, probably the unification scale, and is taken as
input to be explained by a more ambitious theory.

The usual SUSic dark matter does not exist, because the gravitino is
the LSP, and its longitudinal component is strongly coupled.  The
gravitino is relativistic even at today's temperature.  However, the
model contains a long lived penta-baryon, which could be the dark
matter if an appropriate asymmetry in penta-baryon number is
generated in the early universe\cite{bmo} .   One might even hope to
explain the baryon to dark matter ratio of the universe if the
ordinary and penta-baryon asymmetries were generated in the same
physical process.

There is no reason to believe that raising $\lambda$ to its observed
value will make a qualitative change in the electroweak and SUSY
physics. Even in the effective field theory approximation, we were
unable to do precision calculations, because the physics depends on
non-holomoprhic quantities in the strongly coupled Pentagon.  For
realistic values of the parameters one must truly solve the strongly
coupled theory in order to compute precision observables.

The coincidence in scales $10 \Lambda_5^2  \sim \lambda^{1/4} M_P$
may be ``explained" by the anthropic considerations of the appendix,
in which case it is related to the cosmic coincidence between dark
energy and dark matter. In this context it is worth noting that the
nature of Weinberg's galaxy bound changes in this model, because the
properties of dark matter depend on $\lambda$.

It is worth concluding by comparing this model to the SSM.  The
Pentagon model introduces only four new dimensionless parameters,
$g_{S,T}$, $g_{\mu}$ and $g_G$ to the standard model parameter set,
as compared to the $107$ of the SSM.   In the Pentagon model, quark
and lepton flavor are automatically conserved, except for the usual
quark Yukawa couplings and neutrino mass terms. Low energy CP
violation resides only in the usual CKM phase, and the phase of
$h_a$, rather than the myriad new phases of the SSM.  On the face of
it, the model has a strong CP problem, and some mechanism must be
found to solve it.

The Pentagonal mechanism for electroweak symmetry breaking is quite
different from that of the SSM, and may not suffer from the same
fine tuning problems.  The conventional Higgs mass bound is evaded.
Dark matter is a penta-baryon rather than an LSP, and there is no
conventional LSP. The experimental signatures of this kind of dark
matter will be quite different from those of a LSP. Decays of
superpartners of the standard model particles may proceed in a
manner similar to gauge mediated models (though the details of this
remain to be investigated). Note that the gravitino is extremely
light, and causes no cosmological problems (in marked contrast to
most gauge mediated models, and parts of SSM parameter space).

The price for all of these (mostly attractive) differences is an
inability to do precision calculations.   If the Pentagon really
controls the world, high energy theorists and lattice gauge
theorists have their work cut out for them.  It will be important to
find a way of calculating in this strongly coupled theory in order
to make precise comparisons with experiment.  {\it Also, we will
have to verify three crucial dynamical assumptions: that CP is not
violated spontaneously, that the SUSY violating minimum is locally
stable, and that $F_3$, the F term of the independent modulus of the
Pentagon model, is of order $\lambda^{1/4} M_P$.}

In the last section, we showed that all of these attractive features
were retained in a dynamical SUSY breaking model, obtained from the
Pentagon model by a different choice of quantum numbers for the
singlet field, $G$.    This interpretation of the model was more
attractive to the conservative effective field theorist than the
original CSB version, because we did not have to rely on mysterious
assumptions about the behavior of $R$ violating terms induced by
interactions with the cosmological horizon.   Its phenomenological
predictions are similar to those of the CSB model, apart from the
resolution of the axion problem.

From the DSB point of view, key new dynamical features of the model
are the new mechanism for electroweak symmetry breaking, and a
different attitude towards meta-stability of SUSY violating vacua.
In previous work on dynamical SUSY breaking, one always insisted
that the flat space model have no supersymmetric solutions. Here we
instead use the fact that if the c.c. is tuned close to zero at a
meta-stable SUSY violating point, then the model contains no SUSic
solutions. The low energy effective field theory has a stable SUSic
AdS solution, but this is not part of the same quantum theory as the
meta-stable dS minimum. The latter has tunneling amplitudes to a Big
Crunch space-time, but they are so small as to be irrelevant.   A
local observer is super-exponentially\footnote{Super-exponentially
small means an exponential of an inverse power of the c.c. in Planck
units.} more likely to be destroyed by thermally activated processes
in dS space, than by tunneling to a Crunch. Those thermally
activated processes themselves take place on a time scale
super-exponentially longer than the current age of the universe.
These instabilities are phenomenologically irrelevant.  Furthermore,
it is possible that in a true theory of quantum gravity, they are
not instabilities at all, but merely improbable low entropy
fluctuations of a finite system\cite{etinf}.  It may be that even if
the idea of CSB does not survive, these new tricks for constructing
models of dynamical SUSY breaking will come in handy.

The most pressing issue at this point is to work out distinctive
experimental signatures of the Pentagon model.   It certainly
predicts superpartners in the LHC discovery range, and their decays
should be sufficiently different from SSM predictions to distinguish
the models at the LHC.   In particular, the decays of sleptons into
electron plus photon plus missing energy, characteristic of any low
scale SUSY breaking model in which the gravitino is the LSP, should
be a clear signature.  These decays will certainly occur within the
detector in the Pentagon model, because the coupling to the
Goldstino component of the gravitino is quite strong. Although
precise  mass predictions are difficult in this model, it is clear
that sleptons will be quite light.

Finding a Higgs boson above the SSM bounds, in addition to gauginos
at a few hundreds of GeV, would suggest additional degrees of
freedom at the TeV scale, but could not single out the Pentagon as
the culprit. The Pentagon model predicts a rich new strongly
interacting sector, which may be just out of reach of the LHC.  It
is extremely important to search for arguments that some of the
penta-hadrons are anomalously light, which would lead to really
distinctive experimental signatures.

The axion, a very light particle predicted by the Pentagon, might
already be ruled out experimentally.  If not, it is likely to be one
of the more accessible experimental signatures of our model. Finding
it will require higher precision low energy experiments, rather than
the LHC. It is important to estimate its production rate in
accelerators and find its dominant decay modes.  We have presented a
mechanism for raising the axion mass and lowering its coupling, to
make it compatible with both terrestrial experiments and stellar
evolution.  At the moment, we seem to require a $1\%$ fine tuning of
dimensionless couplings to make the model consistent with both these
axion properties and the scale of electroweak symmetry breaking.

Finally, the Pentagon model predicts a distinctive new form of dark
matter, with relatively large magnetic moments.   This may lead to
important observational constraints on the model\cite{caldwell}, or
perhaps a way of discovering evidence for it\footnote{I would like
to thank R. Caldwell for pointing out the strong constraints on the
dipole moments of dark matter particles.}.

\section{Acknowledgments}

I would like to thank S.Thomas for several crucial conversations
which contributed to this work (and for remarks about CP violation
which stimulated the corrections in the current version), and M.
Dine for questions about pseudo-Goldstone bosons, and the relation
of the Pentagon model to dynamical SUSY breaking. N. Seiberg helped
me to understand numerous things about strongly coupled SUSY gauge
theories. I particularly want to thank J.D. Mason for pointing out a
serious error in a previous version of this paper and for several
other comments. I am also grateful to N. Arkani-Hamed for a number
of insightful comments.

This research was supported in part by DOE grant number
DE-FG03-92ER40689.

\section{Appendix 1 - Anthropic Considerations}

Let me begin by stating some opinions about the anthropic principle,
which I view as an antidote to the somewhat thoughtless discussion
of this topic in the high energy theory community and beyond it.
Ever since the advent of inflationary cosmology, we have been faced
with the possibility that theoretical models would describe multiple
regions which remain causally disconnected into the indefinite
future.  If one accepts that possibility then one is immediately led
to consider the possibility that some of the parameters in our low
energy effective Lagrangian may not be derivable from first
principles, but are instead properties of our immediate causal
environment.  Andrei Linde and I \cite{bl} pointed out independently
that in this context an anthropic explanation for the puzzling value
of the cosmological constant might be all we could ever hope for
(Linde had previously discussed the anthropic principle in
inflationary cosmology in\cite{l}.) . Weinberg\cite{wein} did the
crucial calculation which made these considerations quantitative:
{\it other parameters remaining unchanged}, there can be no galaxies
in a universe with positive cosmological constant greater than about
$100$ times its observed value. Refined versions of this
argument\cite{weinvil} claim to find that the observed value is a
``typical" value compatible with life. One does not need to believe
these more precise arguments to realize that the observed value of
the c.c. provides strong, if not compelling evidence for such a
picture.  An important part of the arguments discussed in this
paragraph is the assumption that there is a distribution of values
of the c.c., which is smooth and fairly dense near zero.

There are,in my opinion, two flaws in this general circle of ideas,
the first philosophical, the second experimental.  Part of the often
expressed philosophy of these discussions is that ``using anthropic
arguments in {\it e.g.} eternal inflationary cosmology, is like
using them to understand why we live on earth rather than on the
surface of the sun".  The fallacy here is that, while we can see the
sun and all the other places where we don't live because the
conditions are inhospitable, {\it we cannot, in principle, ever
observe the myriad other universes that are required to exist in
order to explain the c.c. in anthropic terms.}  Therefore,
discussions about them will remain forever in the realm of
meta-physics, and perhaps one day (but certainly not at present)
rigorous mathematics.   Their existence is not subject to
experimental test.   I will refer to models based on such a
metaphysical meta-universe as {\it metaphysical models}.

The second problem is that from a general point of view, and within
the context of most explicit proposals for such a meta-verse, one
would expect the Lagrangian we observe to be the most general one
consistent with our existence, with otherwise random values of the
couplings.  {\it This proposal is ruled out experimentally.  There
are many parameters in the standard model, and its leading
irrelevant corrections, which are much more finely tuned than the
anthropic principle requires}\cite{bdgbdm}.   {\it Anthusiasts},
wanting to have their cake and eat it too, propose to resolve this
by a combination of anthropic reasoning and traditional symmetry
arguments.  But explicit models of a multiverse do not seem to give
any special status to symmetries.   There appear to be vastly more
anthropically allowed ``vacua" without symmetries than with them, as
one might expect from simple considerations of conditional
probability\footnote{Like all such statements, this one should be
taken with a grain of salt.   Investigations of the Landscape of
String Theory - the main proposal for a meta-verse - are at a quite
primitive stage, and almost any claim about it might be wrong.}.

One proposal which avoids these experimental problems is the
``Friendly Landscape" of \cite{adk}.   These authors propose a toy
model for a Landscape in which all but a few parameters have small
fluctuations around a central value.  Furthermore, the parameters
with large fluctuations are precisely those for which we have
unresolved fine tuning problems.   Models like this would be highly
predictive, and one could imagine that the probability distribution
was peaked at symmetric points (since it is determined by
mathematical, rather than anthropic criteria), avoiding the fine
tuning problems for many parameters of a purely anthropic model. It
remains to be seen if such friendly landscapes can be derived from a
more fundamental theory\footnote{Much of the String Landscape
appears to be unfriendly, and one would have to show that the
correct probability distribution favored the friendly regions, if
any.} , and whether they predict the correct central values.

I have gone into some detail in this discussion, in order to
contrast general anthropic arguments with a similar line of argument
which I will use in discussing CSB.   The fundamental claim of CSB
is that there is a countable set of theories of quantum gravity in
an asymptotically de Sitter space with cosmological constant
$\lambda$. The number of quantum states $N$ in each of these
theories is finite, and there is a one to one relation between
$\lambda$ and $N$, with $\lambda$ going to zero as $N$ becomes
infinite. For $\lambda \rightarrow 0$, the theory becomes an ${\cal
N} = 1$ Super Poincare invariant theory with a compact moduli space.
The absence of know examples of such theories suggests that the
limiting theory may be unique, or that the number of possibilities
is much smaller than the number of string vacua that are claimed to
exist in the Landscape. Another important property of these models
is that for finite $N$ they suffer from inherent quantum ambiguities
in what the Hamiltonian and observables are, since the theory does
not contain self consistent measuring devices which can measure
quantum information with arbitrary precision.  So one is led to
consider the small $\lambda$ limit in order to find precise
predictions.

The simplest way to analyze the possible values of $\lambda$ is to
eschew the idea of an underlying meta-physical model, and simply
declare that our theory of the world contains a free, discrete,
dimensionless parameter ($N$).   One then views the Weinberg bound
as simply a gross observational constraint on the value of this
parameter: requiring the theory to have galaxies bounds $\lambda$ by
$$\lambda < c\ Q^3 \rho_0 , $$ where $c$ is a constant of order one,
$Q$ is the amplitude of primordial density fluctuations, and
$\rho_0$ is the dark matter density at the beginning of the matter
dominated era.   In the Pentagon model, $\rho_0$ scales like
$\lambda^{1/4}{M_P \over \Lambda_5} n_0$, where $n_0$ is the density
of P-baryon number.   The primordial ratio of the P-baryon number to
entropy densities (which determines $n_0$), as well as $Q$, are
determined by physics well above the TeV scale and below the Planck
scale. They are likely to be independent of $\lambda$ for small
$\lambda$, and completely fixed by the theory.  This leads to a
bound
$$\lambda^{3/4} < Q^3 {\Lambda_5 \over M_P} n_0 .$$  The right hand side
of this equation can be estimated in the limiting $\lambda = 0$
model.   Of course, if we fix $n_0$ by the phenomenological
requirement that the model reproduce the correct temperature at
which matter domination begins, then this is just Weinberg's bound.
The correct procedure is to assume that $n_0$ is independent of
$\lambda$.   If we assume that it takes on the right observational
value for the observed value of $\lambda$, then we find that galaxy
formation begins at a lower temperature for smaller values of
$\lambda$, and a higher temperature for larger values.  One might
imagine that there are implications for anthropic arguments
following from this observation.  However, we should note that the
temperature of matter radiation equality scales like
$\lambda^{1/16}$ so the constraints are probably not very strong. We
will see much more dramatic effects of changing $\lambda$ in the
microphysics of QCD.

If the c.c. is an input parameter, governing the number of states in
the quantum theory, it is no longer safe to assume that the
probability distribution determining it is flat near $\lambda = 0$.
For example, a flat distribution in the number of states corresponds
to a strong preference for very small $\lambda$. The argument that
we observe a typical value for the c.c. that allows galaxies to
exist is no longer so obvious. A meta-physical model, which
introduces an {\it a priori} preference for large $\lambda$
\cite{bfmholocosm} could solve this problem.  In this model, a
meta-verse consists of a dense black hole fluid (whose coarse
grained description is a flat FRW model with $p = \rho$) in which a
distribution of asymptotically de Sitter bubbles of various sizes
(each bubble consists of exactly one de Sitter horizon volume) match
on to marginally trapped surfaces in the $p = \rho$ geometry (black
holes in the black hole fluid!).   The dS bubbles correspond to
initial conditions of lower entropy than the generic initial
condition which leads to the uniform dense black hole fluid. Thus,
initial conditions which lead to a smaller de Sitter bubble, are
more probable, and we have a preference for the largest cosmological
constant consistent with anthropic bounds.   We would need to know
the functional form of the probability distribution and to decide on
the relevance of refined anthropic considerations\cite{weinvil} in
order to decide whether a cosmological constant of the order we
observe is ``natural".

On the other hand, there might be purely anthropic lower bounds on
$\lambda$ in the Pentagon model. It is extremely interesting that
the qualitative low energy physics and cosmology of our model
changes drastically as soon as $\sqrt{\lambda^{1/4} M_P} \sim 100$
GeV rather than $\sim 3 $ TeV. In particular, QCD does not become
asymptotically free until this rather low scale.   If we assume that
the value of the unified coupling, is independent of $\Lambda$, then
the value of $\alpha_S ( TeV)$ is the same as it is in the real
world. The one loop renormalization group equations for five extra
vectorlike fermions between $1$ TeV and $100$ GeV, then imply that
$\Lambda_{QCD}$ is lowered to a few $MeV$.   There are also changes
to the dimensionless electroweak couplings, but the electroweak
scale is essentially unchanged, since electroweak breaking occurs
for $\lambda = 0$. Note also that the standard model Yukawa
couplings remain essentially unchanged so bare quark masses are the
same. The up and down quark masses will now be of order the
constituent quark mass. Isospin will be strongly broken.

The changes in the electromagnetic and strong interactions,with
fixed weak interaction scale, and quark masses, will have a dramatic
effect on nuclear and stellar physics. I have not worked out the
details, but one can easily imagine that they would lead to an
anthropic lower bound on $\lambda$ in the Pentagon model.

It appears plausible then that the Pentagon model might be the low
energy sector of a model of the world with a single dimensionless
parameter, the cosmological constant in Planck units.   Constraints
on the existence of galaxies and more or less normal stars might
bound this parameter (from both sides) within a few orders of
magnitude of its observed value.

In such a situation, anthropic reasoning is much more attractive
than it is in the context of a landscape. Much of the model,
including the gauge group, is determined in a way which depends very
weakly, or not at all, on $\lambda$.  We do not have to worry about
whether exotic forms of life could exist with different low energy
gauge groups. The parameters which do vary, vary in a calculable
manner, as a function of a single discrete variable.  Very gross
anthropic (really {\it galacto-} or {stellar-} thropic)
considerations bound $\lambda$ within a few orders of magnitude of
its observed value, which is correlated with a variety of other
observables.

\section{Appendix 2}

In the SUSic limit, assuming all Yukawa couplings of order one, $S$
and $T$ vanish and ${\cal M}_t \sim \Lambda_5$.  $H_u H_d \sim
\Lambda_5^2$, follows from the vanishing of $F_S$.  ${\rm tan} \beta
= 1$ follows from the vanishing of the electroweak D-terms.  We can
parametrize the Higgs VEVs by
$$H_u = h_u \pmatrix{1\cr 0},$$
$$H_d = e^{i\phi} \pmatrix{A , B},$$
where $A,B$ and $h_u$ are positive.  If $A \neq 0$ electromagnetism
is spontaneously broken.  In this parametrization, $H_u H_d =
e^{i\phi} h_u B$.   The $F$ term potential constrains $\phi$ and the
product $h_u B$, but there are two components of the physical Higgs
field whose masses are of order $g \Lambda_5$, where $g$ is some
combination of $g_1$ and $g_2$ in the standard model.

Once SUSY is broken, the situation changes.   In general, we may
expect the minimum to be at a place where all $F$ terms are roughly
comparable and of order $\lambda^{1/4} M_P$.  The potential depends
on $S$ in a complicated way, through the Kahler potential  (we still
work in the low energy approximation valid when $\lambda^{1/4} M_P <
M^2$).   Thus, we expect $S$ to get a non-zero VEV.    If it does,
the Higgs potential has a term $$|S|^2 (|H_u|^2 + |H_d|^2) = |S|^2
(A^2 + B^2 + h_u^2) .$$   If $<S> \geq \Lambda_5$,  as we may expect
when we approach the observed value of the c.c.,  then the
combination of this term and the $|F_S|^2$ term in the potential,
gives mass $ \sim \Lambda_5$ to all components of the Higgs field.
Note also that this potential favors $A = 0$, preserving
electromagnetism, just like the electroweak D term potential.

The potential for the Higgs fields $h_u $ and $B = h_d$ will have
the form
$$a^2 \Lambda_5^2 (h_u^2 + h_d^2)  + (g_S \Lambda_5^2 - g_{\mu} h_u
h_d)^2 .$$  $a$ is a parameter giving the VEV of $S$ in $\Lambda_5$
units.  This is minimized at $h_u = h_d = {1\over\sqrt{2}} v$ where

$$a^2 \Lambda_5^2 +{g_{\mu} \over 2}({g_{\mu} \over 2} v^2 - g_S
\Lambda_5^2) = 0.$$   The value of $v^2$ is given by the difference
of two positive functions of the couplings and can be made smaller
than $\Lambda_5$ by fine tuning.   To get $v = 250 GeV$, with
$\Lambda_5 = 3 TeV$ we require a fine tuning of about $1\%$.




  %




\end{document}